\documentclass{elsarticle}
\newtheorem{thm}{Theorem}[section]
\newtheorem{prop}{Proposition}[section]
\newtheorem{defn}{Definition}[section]
\newtheorem{lemma}{Lemma}[section]

\newtheorem{cor}{Corollary}[section]
\newtheorem{examo}{Example}[section]
\newtheorem{exams}{Examples}[section]

\newcommand{\set}[2]{\{#1\mid#2\}}

\newcommand{\ua}{\mathord{\uparrow}}
\newcommand{\da}{\mathord{\downarrow}}
\usepackage{amsmath}
\usepackage{amssymb}
\usepackage{hyperref}
\usepackage{color}
\numberwithin{equation}{section}
\usepackage{cleveref}
\makeatletter
\def\ps@pprintTitle{%
  \let\@oddhead\@empty
  \let\@evenhead\@empty
  \def\@oddfoot{\reset@font\hfil\thepage\hfil}
  \let\@evenfoot\@oddfoot
}
\makeatother
\begin{document}
\begin{frontmatter}
\title{The categorical equivalence between disjunctive sequent calculi and algebraic L-domains}
\author[1]{Longchun Wang}
\ead{longchunw@163.com}
\address[1]{School of Mathematics, Hunan University, Changsha, Hunan, 410082, China}

\author[1]{Qingguo Li\corref{a1}}
\cortext[a1]{Corresponding author.}
\ead{liqingguoli@aliyun.com}
\address[2]{School of Mathematical Sciences, Qufu Normal University, Qufu, Shandong, 273165, China}

 \begin{abstract}

This paper establishes a  purely syntactic representation for the category of algebraic $L$-domains with Scott-continuous functions as morphisms. The central
tool used here is the notion of  logical states,  which builds a bridge between  disjunctive sequent calculi  and algebraic $L$-domains. To capture Scott-continuous functions between algebraic $L$-domains, the notion of consequence relations between disjunctive sequent calculi is also introduced. It is shown that the category of disjunctive sequent calculi with consequence relations as morphisms is categorical equivalent to that of algebraic $L$-domains with Scott-continuous functions as morphisms.
\end{abstract}

\end{frontmatter}
\section{Introduction}
Domains with Scott-continuous functions, introduced by Scott and Strachey in the 1970s~(see\cite{b1,b2}), form a foundational denotational semantics for functional programming languages in computer science.  Apart from the  extensive applications in computer
science, domains are the important
objects discussed in mathematics and have  overlaps with order theory, topology, logic, formal concept analysis and category theory~(see\cite{b3,b4,b5,b6,b7,b8}).
The connection between logic and domains is intricate and well-known.
One aspect of the connection between logic and domains is that  domains can be presented by logical languages, and
this connection has demonstrated by Scott's information systems~\cite{b9} and Amramsky's domain logic~\cite{b10, b11}.

Scott first demonstrated the  possibility of logic syntactic representation for Scott domains by the notion of   information systems~\cite{b9}.  An information system is a triple~$(X,Con,\vdash)$, which is introduced and used to substitute for the  domain-theoretical approach to the semantics of functional programming languages. Here $X$ is a set of atomic formulae and~$Con$ is  a   collection of finite subsets of atomic formulae.   $\vdash$ is an entailment relation from $Con$ to $X$ and tells us what atomic formulae  can be entailed from a member of $Con$. A Scott domain can be formed by taking those subsets of $X$ closed under the entailment relation~$\vdash$  as the points, and this kind of construction of domains is easy to understand for computer scientists~\cite{b9}.
 In \cite{b12}, the category of  Scott domains with Scott-continuous functions as morphisms was also shown to be equivalent to that of Scott's information systems with approximable  mappings as morphisms. Now, a lots of information systems have been introduced and used to
  represent various domains (see\cite{b13, b14, b15, b16}).

In a logical semantics theoretical development, Abramsky  devised a complete logic description for Scott domains  and show how the domain logic can be usefully employed in denotational semantics \cite{b10}.  Abramsky's domain logic, as well as Scott's information system, is made by extracting an appropriate logical language from the category of Scott domains with Scott-continuous functions as morphisms. However,  it allows logical formulae can be combined by
connectives,
  and it has more functional properties than information systems from a  perspective of program logic interpretation.  His work was technically demanding but rewarding, recognised in the first LICS "test of time" award.
   Many researchers  have  tried to apply Abramsky's program to other classes of domains (see \cite{b17,b18,b19,b20,wang2}).  The most related work for us is due to  Chen and Jung's paper \cite{b21}, in which  they developed a framework of  disjunctive propositional logic and provided a logic algebraic representation for the category of algebraic L-domain with stable functions as morphisms using its Lindenbaum algebra.

This paper aims to introduce a logical syntactic  representation for algebraic $L$-domains from the categorical viewpoint. Although the investigation  is  inspired by the work of  Chen and Jung \cite{b21}, our approach  differs greatly.   We not only provide a purely syntactic representation of algebraic $L$-domains, but also we are particularly interest in the suitable logical representation of Scott-continuous functions between algebraic $L$-domains.

 In  \Cref{3.1}, based on the analysis of some properties of contradiction, tautology, satisfiable formula and conjunction  in a disjunctive propositional logic,  the concept of logical states in a disjunctive propositional logic is proposed.
A logical state is  a  pattern  of   formulae that satisfies two conditions, and
  there are two methods for generating  logical states.
 In \Cref{3.2}, it is shown that the collection of all logical states of a disjunctive propositional logic ordered by  set inclusion forms an algebraic $L$-domain. Conversely,  each algebraic $L$-domain can be generated in this way, up to isomorphism. Then a new representation theorem for algebraic L-domains is obtained: instead of making use of  the Lindenbaum algebra of the disjunctive propositional logic, the whole process simply employs the satisfies formulae of disjunctive propositional logic.

There is another difference from the work by Chen and Jung \cite{b21}:  the morphisms between algebraic L-domains that they discussed was stable functions, which are fairly strict about the Scott-continuous functions  between algebraic L-domains.
It is well known that Scott-continuous functions are typically used as morphisms between algebraic L-domains to form a cartesian closed category $\mathbf{ALD}$.
 In \Cref{4}, we  introduce  a consequence relation between expressive disjunctive sequent calculi, which has some similar features to a consequence relation between multi lingual sequent calculi introduced in~\cite{b17}. We show that there is a one to one correspondence between consequence relations on  expressive disjunctive sequent calculi and Scott-continuous functions on algebraic $L$-domains.
   Finally, we establish a logic category which is  cartesian closed and equivalent to $\mathbf{ALD}$. This result provides a logical characterization for denotational semantics of functional programming languages and a potential valuable application in theoretical computer science.

   \section{Preliminaries}
We first  recall some basic definitions and notations which will be  used in this article.
\subsection{Order and domain theoretical notations}
 Our order and domain theoretical notation and terminology are  standard,
 most of them come from~\cite{b22,b23,b24}.

 Let~$P$ be a poset. If  $P$ has a least element~$\bot$, then $P$ is called \emph{pointed}.
 A nonempty subset $D$ of a $ P$ is said to be \emph{directed} if every pair of elements of $D$ has an upper bound in~$D$.  We use $\da X$ to denote the down set $$\set{d\in P}{(\exists x\in X)d\leq x},$$ where $X$ is a subset of $P$. Similarly, we write $\ua X$  for the upper set $$\set{d\in P}{(\exists x\in X)x\leq d}.$$ If $X$ is a singleton~$\{x\}$, then we just write $\da x$ or $\ua x$.  $X$ is a \emph{pairwise inconsistent} subset of  $P$ if $\ua x\cap\ua y=\emptyset$ for all $x\neq y\in X$.
   $P$ is said to be a \emph{complete lattice} if each subset $X$ of it has a supremum~$\mathrm{sup}X$.
  A \emph{dcpo} $P$ is a poset in which every directed
subset $D$ has a supremum~$\mathrm{sup}D$.

Let $P$ be a dcpo and $x\in P$.  If for all directed subset $D$ of $P$ the relation $x\leq \mathrm{sup}D$ always implies the existence of some $d\in D$ with $x\leq d$, then $x$ is called a \emph{compact element} of $P$.  We write $K(P)$ for the set of compact elements of $P$ and write $K^*(P)$ for $K(P)-\{\bot\}$.

\begin{defn} \label{d2.1} \cite{b24}

\begin{enumerate}\label{d2.1}
\item [\emph{(1)}]
A pointed dcpo~$P$ is called  an algebraic domain if every element $x$ of $P$ is the directed supremum of the compact elements
below $x$.
\item [\emph{(2)}]
An algebraic $L$-domain $P$ is any algebraic domain in which for all element $x\in P$, $\da x$ is a complete lattice.
\end{enumerate}
\end{defn}

Noting that an algebraic $L$-domain $P$ satisfies the following property: For all nonempty finite subset $\{x_1,x_2,\cdots,x_n\}$ of $K(P)$ with an upper bound in $P$, there exists a unique nonempty  pairwise inconsistent subset $A\subseteq K(P)$ such that $\ua x_1\cap\ua x_2\cap\cdots\cap\ua x_n=\bigcup_{a\in A}\ua a$. In fact, the set $A$ just is the  minimal upper bounds of $\{x_1,x_2,\cdots,x_n\}$.

\begin{defn}\label{d2.2} \cite{b24}

 Let $P$ and $Q$ be algebraic $L$-domains. A function~$f:P\rightarrow Q$ is Scott-continuous if and only if for all directed subset $D$ of $P$, $f(\mathrm{sup}D)=\mathrm{sup}\set{f(x)}{ x\in D}.$
\end{defn}


\subsection{Disjunctive propositional logic}

Based on  the work by Chen on stable domains \cite{b25} and the work by Zhang on disjunctive information systems \cite{b26},  a disjunctive propositional logic was introduced in \cite[Definition~2.1]{b21}.

 The connectives of a disjunctive propositional logic consists of  two unitary constant connectives T and F, a binary conjunctive connective $\wedge$ and an arbitrary, but provably disjoint,
disjunctive connective~$\dot{\bigvee}$. Starting with a set of atomic formulae and a set of atomic disjointness assumptions, disjunctive formulae are built.
 And in a disjunctive propositional logic, a sequent is an object $\Gamma\vdash \varphi$, in which $\Gamma$ is a finite set of formulae and $\varphi$ is a single formula. In the sequel, we use $A\sqsubseteq B$ to denote $A$ is a finite subset of $B$.

\begin{defn}\label{d2.3} \emph{(\cite{b21})}

Let $P$ be a set, every element of which is  called an atomic  formula. Likewise, let $\mathcal{A}_P$
be a set of sequents of the form $p_1,p_2,\ldots,p_n\vdash F$ where the $p_i$ are atomic formulae, and $\emph{F}$ is the syntactic constant
for ``false''. Each element of $\mathcal{A}_P$ is called an atomic disjointness assumptions, and the pair $(P,\mathcal{A}_P)$ is called a disjunctive basis.

The set $\mathcal{L}(P)$ of  formulae, and the set $\mathbf{T}(P)$ of valid sequents
are generated by mutual transfinite induction by the following rules:
\begin{enumerate}
\setlength{\itemsep}{0cm}
  \item [$\bullet$]  Disjunctive formulae
  \begin{align*}
  &\left(\emph{At}\right)\cfrac{\phi\in P}{\phi\in\mathcal{L}(P)}
&&\left (\emph{Const}\right)\cfrac{ }{\emph{T,F}\in\mathcal{L}(P)}\\
&\left(\emph{Conj}\right)\cfrac{\phi,\psi\in\mathcal{L}(P) }{\phi\wedge\psi\in\mathcal{L}(P)}&&\left(\emph{Disj}\right)\cfrac{\phi_{i}\in\mathcal{L}
(P)(all~i\in I)~~~~~\phi_{i},\phi_{j}\vdash \emph{F} (all~i\neq j\in I) }{\dot{\bigvee}_{i\in I}\phi_{i}\in\mathcal{L}(P)}
\end{align*}

  \item [$\bullet$] Valid sequents
\begin{align*}
&\left(\emph{Ax}\right)\cfrac{(\Gamma\vdash \emph{F})\in \mathcal{A}_P}{\Gamma\vdash \emph{F}}&&
\left (\emph{Id}\right)\cfrac{\phi\in \mathcal{L}(P) }{\phi\vdash \phi}\\
&\left(\emph{Lwk}\right)\cfrac{\Gamma\vdash \psi~~~~~~~\phi\in\mathcal{L}(P) }{\Gamma,\phi\vdash \psi}&&
\left (\emph{Cut}\right)\cfrac{\Gamma\vdash\phi~~~~~~\Delta,\phi\vdash \psi}{\Gamma,\Delta\vdash\psi}\\
&\left(\emph{LF}\right)\cfrac{\phi\in\mathcal{L}(P)}{\emph{F}\vdash\phi}
&&
\left (\emph{RT}\right)\cfrac{~~~~~~~~~~}{\vdash \emph{T}}\\
&\left(\emph{L}\wedge\right)\cfrac{\Gamma,\phi,\psi\vdash \theta }{\Gamma,\phi\wedge\psi\vdash \theta}
&&
\left (\emph{R}\wedge\right)\cfrac{\Gamma\vdash\phi~~~~~~~~~\Delta\vdash \psi}{\Gamma,\Delta\vdash\phi\wedge\psi}
\end{align*}
$$\left(\emph{L}\dot{\vee}\right)\cfrac{\Gamma,\phi_{i}\vdash \theta (all~i\in I)~~~~~~~~~~~\phi_{i},\phi_{j}\vdash \emph{F} (all~i\neq j\in I) }{\Gamma,\dot{\bigvee}_{i\in I}\phi_{i}\vdash \theta}~~~~~~~~~~~~
$$

$$\left(\emph{R}\dot{\vee}\right)\cfrac{\Gamma\vdash\phi_{i_0} (some~i_0\in I)~~~~~~~~~~~\phi_{i},\phi_{j}\vdash \emph{F} (all~i\neq j\in I) }{\Gamma\vdash\dot{\bigvee}_{i\in I}\phi_{i}}.~~~~~~~~
$$
\end{enumerate}
\end{defn}

  The proof system of a  disjunctive propositional logic is sound and complete  with respect to its Lindenbaum algebra~\cite{b21}.

    With respect to the  disjunctive propositional logic proposed in Definition~\ref{d2.3}, we can define a  binary relation $\Vdash$ between the finite subsets of $\mathcal{L}(P)$ and the set $\mathcal{L}(P)$ as following:
   \begin{equation}
   (\Gamma,\varphi)\in \Vdash \mathrm{~if~and~only~if~the~sequent~}\Gamma\vdash\varphi\mathrm{~is~valid}.
   \end{equation}
   In other words, a valid sequent $\Gamma\vdash\varphi$ and $(\Gamma,\varphi)\in \Vdash$ are mutual determined. Then  the disjunctive propositional logic   can be seen as  a pair~$(\mathcal{L}(P),\Vdash)$ such that~$\mathcal{L}(P)$ and $\Vdash$ are closed under the rules of disjunctive formulae and valid sequents. In this article no differentiation between $\Vdash$ and  $\vdash$ is made  for convenience, and the pair  $(\mathcal{L}(P),\vdash)$ is called  a disjunctive sequent calculus.

\begin{prop}\label{prop2.1}~\emph{(\cite{b21})}

Let~$(\mathcal{L}(P),\vdash)$ be  a disjunctive sequent calculus. Then the following statements hold.
\begin{enumerate}
\item [\emph{(1)}] $\Gamma, \varphi,\psi\vdash\phi$ is a valid sequent if and only if $\Gamma, \varphi\wedge\psi\vdash\phi$ is a valid sequent.
\item [\emph{(2)}] $\Gamma\vdash\varphi$ and $\Gamma\vdash\psi$ are valid sequents if and only if $\Gamma\vdash\varphi\wedge\psi$ is a valid sequent.
\item [\emph{(3)}] Assuming $\phi_i,\phi_j\vdash\emph{F}$ are valid sequents for all $i\neq j\in I$, then $\Gamma,\phi_i\vdash\theta$ are valid sequents if and only if $\Gamma,\dot{\bigvee}_{i\in I}\phi_i\vdash\theta$ is a valid sequent.
\end{enumerate}
\end{prop}

   In the scheme of the Lindenbaum algebra of a disjunctive propositional logic, Chen studies the Stone duality and a logic algebraic representation for the category of algebraic L-domain
 with  stable functions as morphisms \cite{b25}.

Instead of using the Linderbaum algebra of a disjunctive propositional logic, we turn to provide a logical syntactic representation of algebraic $L$-domains using the proof system of a disjunctive propositional logic. Moreover, in viewpoint of category, the morphisms between algebraic $L$-domains what we focus on are Scott-continuous functions which are typically employed in domain theory.

\section{Logical representation}
In this section, we show  how  to use the disjunctive sequent calculi to
  represent algebraic L-domains. We unfold completely in the  class $\mathcal{L}(P)$ of   a disjunctive sequent calculus $(\mathcal{L}(P),\vdash)$, without  assistance of  any object except valid sequents.
\subsection{Logical states}\label{3.1}

We begin by introducing  some common  terms in a disjunctive sequent calculus.

\begin{defn}\label{d3.1}
Let~$(\mathcal{L}(P),\vdash)$ be a disjunctive sequent calculus.
 \begin{enumerate}
\item [\emph{(1)}] A formula $\varphi$ is said to be a tautology if the sequent $\emph{T}\vdash \varphi$ is  valid.
\item [\emph{(2)}] A formula  $\varphi$ is said to be a contradiction if the sequent $\varphi\vdash\emph{F}$ is  valid.
  \item [\emph{(3)}] A formula  $\varphi$ is said to be  satisfiable  if it is neither a tautology nor a contradiction.
  \end{enumerate}
\end{defn}

   Given an atomic formula $p$ of  a disjunctive sequent calculus  $(\mathcal{L}(P),\vdash)$, it is easy to see that  the formula  $\mathrm{F}\wedge p$ is  a contradiction  and  the formula $\mathrm{T}\dot{\bigvee} (\mathrm{F}\wedge p)$ is a tautology. We denote by Tau$(P)$ and Cont$(P)$
the set of all tautologies and the set of all contradictions, respectively.

\begin{defn}\label{d3.2}

Let~$(\mathcal{L}(P),\vdash)$ be a disjunctive sequent calculus and $\varphi,\psi\in \mathcal{L}(P)$.  If both $\varphi\vdash\psi$ and $\psi\vdash\varphi$ are valid sequents, then $\varphi$ and $\psi$ are said to be logically equivalent.
\end{defn}

Noting that in~\cite{b21},  two logically equivalent formulae  are called interderivable.

\begin{defn}\label{d3.3}

Let~$(\mathcal{L}(P),\vdash)$ be a disjunctive sequent calculus.
 \begin{enumerate}
\item [\emph{(1)}]

A conjunction is a satisfiable formula that built up from atomic formulae only by conjunctive connectives.
\item [\emph{(2)}]
 A flat formula is a  satisfiable formula that has the form $\dot{\bigvee}_{i\in I}\mu_i$, where  $\mu_i$ are   conjunctions  and $\mu_i,\mu_j\vdash \emph{F}$ are valid for all $i\neq j\in I$.
 \end{enumerate}
\end{defn}

\begin{prop}\label{p3.1}~\emph{(\cite{b21})}

 Every  satisfiable formula in a disjunctive sequent calculus~$(\mathcal{L}(P),\vdash)$ is logically equivalent to  a flat formula $\dot{\bigvee}_{i\in I}\mu_i$, where $I$ is a nonempty set and $\mu_i$ are conjunctions for all $i\in I$.

\end{prop}

  We denote by $\mathcal{N}(P)$  the set of all  flat formulae in a disjunctive sequent calculus~$(\mathcal{L}(P),\vdash)$.
 Next we will present the notion of a logical state, which constructs an important bilateral link  between disjunctive sequent calculi and algebraic $L$-domains. The name of a logical state borrows from the notion of a ``state'' in  information systems(see~\cite{b13, b14}).

\begin{defn}\label{d3.4}

Let~$(\mathcal{L}(P),\vdash)$ be  a disjunctive sequent calculus.
 A  logical state of $(\mathcal{L}(P),\vdash)$ is a nonempty proper subset $S$ of $\mathcal{L}(P)$ such that the following  conditions hold:
\begin{enumerate}
\item [{\bf(S1)}] If $ \dot{\bigvee}_{i\in I}\mu_i\in S\cap \mathcal{N}(P)$, then there exists some $i_0\in I$ such that $\mu_{i_0}\in S,$.
 \item [{\bf(S2)}] $S[\vdash]\subseteq S$,
 where
\begin{equation}\label{eq3.1}
X[\vdash]=\set{\varphi\in \mathcal{L}(P)} {(\exists \Gamma\sqsubseteq X)\Gamma\vdash \varphi \in{ \mathbf{T}}(P)}
 \end{equation}
 for all  $X\subseteq \mathcal{L}(P)$.
 \end{enumerate}
\end{defn}

We denote  by $|(\mathcal{L}(P),\vdash)|$ the set of all  logical states of a disjunctive sequent calculus $(\mathcal{L}(P),\vdash)$.

Let $\varphi$ be a  satisfiable formula in $S$.  As we have seen in Proposition \ref{p3.1} there is a flat formula $\dot{\bigvee}_{i\in I}\mu_i$ logically equivalent to the formula $\varphi$.   Condition~(S2) suggests that  the formula $\dot{\bigvee}_{i\in I}\mu_i$ is  a member of $S$. Then by condition (S1) we have some $i_0\in I$ such that the conjunction $\mu_{i_0}$ is also a member of $S$. Moreover, condition~(S2) indicates that a logical state $S$ is closed under $\vdash$. The following proposition shows that $S$ is  closed under $\wedge$.

\begin{prop} \label{p3.2}

Let $S$ be a logical state of a disjunctive sequent calculus $(\mathcal{L}(P),\vdash)$. Then the following statements hold.
\begin{enumerate}
\item [\emph{(1)}] If $\varphi,\psi \in S$, then $\varphi\wedge\psi\in S$.
\item [\emph{(2)}] The constant $\emph{F}$ does not belong to $ S$.
\item [\emph{(3)}]  $S=S[\vdash]$.
\item [\emph{(4)}] If $\varphi\in S$ and $\varphi\wedge \psi\vdash \emph{F}$ is valid, then $\psi\notin S$.
\item [\emph{(5)}] If  $X$  is a subset of $S$, then
 \begin{equation}\label{eq3.2}
 [X]_S=\bigcap\set{W\in|(\mathcal{L}(P),\vdash)|}{X\subseteq W\subseteq S}
  \end{equation}
  is also a logical state.
\end{enumerate}
\end{prop}
$\mathbf{Proof.}$

 (1) By Definition~\ref{d2.3} and Proposition \ref{prop2.1}, $\varphi,\psi\vdash\varphi\wedge\psi$ is a valid sequent.  Then according to \Cref{eq3.1} and condition~(S2), we have $\varphi\wedge\psi\in S$.

(2) Suppose not, then $\text{F}\in S$. By the rule (LF),  $\text{F}\vdash\psi$ is valid for any $\psi\in\mathcal{L}(P)$. Thus $\mathcal{L}(P)\subseteq S\subseteq \mathcal{L}(P)$. This contradicts the fact that $S$ is a proper subset of $\mathcal{L}(P)$.

(3) $S[\vdash]\subseteq S$ just is an application of condition~$(S2)$, and $S\subseteq S[\vdash]$ follows immediately by \Cref{eq3.1} and the rule~(Id).

(4) This follows from parts~(2) and (3).

(5) According to \Cref{eq3.1} and the rule (RT), it follows that the constant $\mathrm{T}$ is a member of $S$. This implies that $[X]_S$ is nonempty.

Assume that $\dot{\bigvee}_{i\in I}\mu_i\in [X]_S\cap \mathcal{N}(P)$. Then by \Cref{eq3.2}, we have $\dot{\bigvee}_{i\in I}\mu_i\in W$ for all $W\in |(\mathcal{L}(P),\vdash))|$ with $X\subseteq W\subseteq S$. Since $W$ is a logical state, there exists some $i_W\in I$ such that $\mu_{i_W}\in W\subseteq S$.  Thus $\set{\mu_{i_W}}{W\in |(\mathcal{L}(P),\vdash)|, X\subseteq W\subseteq S}$ is a subset of $S$. Note that $\mu_i,\mu_j\vdash \text{F}$ is valid for all $i\neq j\in I$ and  $S$ is a logical state, it is not difficult to show that the set  $\set{\mu_{i_W}}{W\in |(\mathcal{L}(P),\vdash)|, X\subseteq W\subseteq S}$ is a singleton. This implies that there exists some $i_0\in I$ such that $\mu_{i_0}\in[X]_S$. Therefore, we have shown that $[X]_S$ satisfies condition (S1).

For condition~(S2), assume that $\varphi\in[X]_S[\vdash]$. Then there exists some $\Gamma\sqsubseteq [X]_S$ such that $\Gamma\vdash \varphi$. Thus $\Gamma \sqsubseteq W$ for any $W\in|(\mathcal{L}(P),\vdash)|$ with $X\subseteq W\subseteq S$. Since $W$ is a logical state and $\Gamma\vdash \varphi$, it follows that $\varphi\in W$, and therefore $\varphi\in[\Gamma]_S$.
\hfill$\square$


\begin{prop} \label{p3.3}

Let $(\mathcal{L}(P),\vdash)$ be a disjunctive sequent calculus. Then the following statements hold.
\begin{enumerate}
\item [\emph{(1)}] \emph{Tau}$(P)$ is a logical state but \emph{Cont}$(P)$ is not.

 \item [\emph{(2)}] The union of a directed subset of logical states is a logical state.
 \end{enumerate}
\end{prop}
$\mathbf{Proof.}$

(1)  Since Tau$(P)\cap  \mathcal{N}(P)=\emptyset$,      Tau$(P)$  naturally fulfills  condition~(S2).  By the  definitions of a tautology and Tau$(P)[\vdash]$, it is easy to see that Tau$(P)[\vdash]\subseteq \text{Tau}(P)$.  Consequently, Tau$(P)$ is a logical state.

 According to part~(3) of Proposition~\ref{p3.2}, it follows that Cont$(P)$ is not a logical state .

(2) For a directed  set of logical states $\set{S_i}{i\in I}$, set $S=\bigcup\set{S_i}{i\in I}$. We show that $S$ is a logical state by checking that $S$ satisfies conditions $(S1)$ and $(S2)$.

  To prove condition $(S1)$, let $ \dot{\bigvee}_{j\in J}\mu_j\in S\cap \mathcal{N}(P)$. Then $ \dot{\bigvee}_{j\in J}\mu_j\in S_{i}\cap \mathcal{N}(P)$ for some $i\in I$. Thus there exists some $ j_{i}\in J$ such that $\mu_{j_i}\in S_i\subseteq S$  because $S_i$ is a logical state.

  For condition~$(S2)$, let $\varphi\in S[\vdash]$. By equation~(\ref{eq3.1}), there exists some $\Gamma \sqsubseteq S$ such that $\Gamma\vdash\varphi$. From the fact that $\Gamma\sqsubseteq S$ and the set $\set{S_i}{i\in I}$ is  directed,  it follows that $\Gamma \sqsubseteq S_{i_0}$ for some $i_0\in I$. Since $S_{i_0}$ is a logical state,  we have $\varphi\in S_{i_0}\subseteq S$. Therefore, $S[\vdash]\subseteq S$.
 \hfill$\square$

 Part (5) of Proposition~\ref{p3.2} and
part (2) of Proposition~\ref{p3.3}  present two methods for deriving new logical states. Part (1) of Proposition~\ref{p3.3} shows there is a trivial logical state~{Tau}$(P)$ for all disjunctive sequent calculus~$(\mathcal{L}(P),\vdash)$. We now show there are enough many nontrivial logical states. To this end, we need a further definition.

  \begin{defn}\label{d3.5}

 Let $(\mathcal{L}(P),\vdash)$ be a disjunctive sequent calculus.
\begin{enumerate}
\item[{(1)}]
A conjunction  $\mu$ is said to be irreducible if,  whenever $\mu\vdash\dot{\bigvee}_{i\in I}\phi_{i}$ is valid, where $\phi_{i},\phi_{j}\vdash \emph{F}$ are valid for all $i\neq j\in I$, then $\mu\vdash \phi_{i_0}$ is valid for some $i_0\in I$.
\item[{(2)}]
A flat formula $\dot{\bigvee}_{i\in I}\mu_{i}$  is irreducible if each conjunction $\mu_{i}$ is irreducible.
\end{enumerate}

\end{defn}

For a disjunctive sequent calculus $(\mathcal{L}(P),\vdash)$,
we denote by $\mathcal{IC}(P)$ the set of all irreducible conjunctions.
Obviously, each irreducible conjunction is an irreducible flat formula.

\begin{prop}\label{p3.7}

If $\mu$ is an irreducible conjunction in a disjunctive sequent calculus $(\mathcal{L}(P),\vdash)$, then $\{\mu\}[\vdash]$ is a logical state.
\end{prop}
$\mathbf{Proof.}$

 Assume that $\dot{\bigvee}_{i\in I}\mu_{i}$ is a flat formula in $\{\mu\}[\vdash]$. Then $\mu\vdash \dot{\bigvee}_{i\in I}\mu_{i}$. Since $\mu$ is an irreducible conjunction, there exists some $i_0\in I$ such that $\mu\vdash \mu_{i_0}$. Therefore, $\mu_{i_0}\in \{\mu\}[\vdash] $.

 Using the rule (Cut), it is trivial to check that the set $(\{\mu\}[\vdash])[\vdash]$ is a subset of $\{\mu\}[\vdash]$.
\hfill$\square$

\subsection{Representation of algebraic $L$-domains}\label{3.2}
 In this subsection, we establish a  satisfactory correspondence between disjunctive sequent calculi and algebraic $L$-domains.
  Before stating and proving our representation theorem, we
require  further   lemmas which are of interest
in their own right.

\begin{lemma}\label{lm3.1}

Let $(\mathcal{L}(P),\vdash)$ be a disjunctive sequent calculus. Then the family $|(\mathcal{L}(P),\vdash)|$ forms a point dcpo under the set inclusion.
\end{lemma}
$\mathbf{Proof.}$

 It is easy to see that Tau$(P)$ is the least element of $(|(\mathcal{L}(P),\vdash)|,\subseteq)$.

 Let $\set{S_i}{i\in I}$ be a directed set of $(|(\mathcal{L}(P),\vdash)|,\subseteq)$. By part (2) of Proposition~\ref{p3.3}, the union $\bigcup \set{S_i}{i\in I}$ is a logical state. Then the supremum of the set $\set{S_i}{i\in I}$ exists in $(|(\mathcal{L}(P),\vdash)|,\subseteq)$, and $$\mathrm{sup} \set{S_i}{i\in I}=\bigcup \set{S_i}{i\in I}.$$   As a result,
 $(|(\mathcal{L}(P),\vdash)|,\subseteq)$ forms a pointed dcpo.
\hfill$\square$

\begin{lemma}

 For any logical state $S$ and finite subset $\Gamma$ of $S$, the set $[\Gamma]_S$ defined by equation~(\ref{eq3.2}) is a compact element of the dcpo $(|(\mathcal{L}(P),\vdash)|,\subseteq)$.
 \end{lemma}
 $\mathbf{Proof.}$

 Let $\set{S_i}{i\in I}$ be a directed subset in the dcpo $(|(\mathcal{L}(P),\vdash)|,\subseteq)$ with $[\Gamma]_S\subseteq \bigcup\set{S_i}{i\in I}$. Since $\bigcup\set{S_i}{i\in I}$ is a logical state, it is evident that $[\Gamma]_S=[\Gamma]_{\bigcup\set{S_i}{i\in I}}$. Because  $\Gamma\sqsubseteq \bigcup\set{S_i}{i\in I}$, there exists some $i_0\in I$ such that $\Gamma\subseteq S_{i_0}$. Whence, $[\Gamma]_S=[\Gamma]_{\bigcup\set{S_i}{i\in I}}\subseteq S_{i_0}$, which implies that  $[\Gamma]_S$  is a compact element of the dcpo $(|(\mathcal{L}(P),\vdash)|,\subseteq)$.
\hfill$\square$
\begin{lemma}

If $S$ is a logical state, then the set $\set{[\Gamma]_S}{ \Gamma\sqsubseteq S}$ is directed and $S$ is its union.
\end{lemma}
$\mathbf{Proof.}$

Let $\Gamma_1,\Gamma_2\sqsubseteq S$. Then $[\Gamma_1\cup \Gamma_2]_S\in \set{[\Gamma]_S}{\Gamma\sqsubseteq S}$. Obviously, $[\Gamma_1]_S,[\Gamma_2]_S\subseteq[\Gamma_1\cup \Gamma_2]_S$, which implies the set $\set{[\Gamma]_S}{\Gamma\sqsubseteq S}$ is directed.

 We next show $S=\bigcup\set{[\Gamma]_S}{\Gamma\sqsubseteq S}$. It is clear that $\bigcup\set{[\Gamma]_S}{\Gamma\sqsubseteq S}\subseteq S$, since $[\Gamma]_S\subseteq S$ for any $\Gamma\sqsubseteq S$. Conversely, suppose that $\varphi\in S$. Then $\varphi\in[\{\varphi\}]_S\subseteq\set{[\Gamma]_S}{\Gamma\sqsubseteq S}$. This yields that $S\subseteq \bigcup\set{[\Gamma]_S}{\Gamma\sqsubseteq S}$.
\hfill$\square$

\begin{thm}\label{t3.1}

Let $(\mathcal{L}(P),\vdash)$ be a disjunctive sequent calculus. Then $(|(\mathcal{L}(P),\vdash)|,\subseteq)$ is an algebraic $L$-domain.
\end{thm}
$\mathbf{Proof.}$

 According to the above lemmas, it follows that $(|(\mathcal{L}(P),\vdash)|,\subseteq)$ forms an algebraic domain. To complete the proof, it suffices to show that for all logical state $S$, the principle ideal $\da_{\subseteq} S$ is a complete lattice. For this, suppose that $\mathcal{T}$ is a subset of $\da_{\subseteq} S$. Just as the proof of part (5) of Proposition~\ref{p3.2}, we can verify the intersection $\bigcap_{T\in\mathcal{T}}T$  is also a logical state. Clearly, $\bigcap_{T\in\mathcal{T}}T\in \da_{\subseteq} S$, and therefore, $\bigcap_{T\in\mathcal{T}}T$ is an infimum of $\mathcal{T}$ in $\da_{\subseteq} S$.
\hfill$\square$

Theorem \ref{t3.1} tells us that each disjunctive sequent calculus associates an algebraic $L$-domain. We now consider the inverse direction.

 \begin{defn}\label{dd3.6}

 Let $(D,\leq)$ be an algebraic L-domain. A subset $U$ of $D$ is said to be decomposable if it is a down set $\ua A$ with $A$ is a pairwise inconsistent subset of $K(D)$.
 \end{defn}

 The notation $\mathcal{U}(D)$ will denote the collection of all decomposable subsets of $D$.
It is obvious that every decomposable set is Scott  open and  $\mathcal{U}(D)$
is closed under finite intersections~$\cap$ and arbitrary disjoint unions~$\dot{\bigcup}$, since $(D,\leq)$ is an algebraic L-domains.

Thus we can make a concrete disjunctive sequent calculus  as follows.

\begin{thm}\label{p3.5}

  Associated with a given algebraic $L$-domain $(D,\leq)$,  a disjunctive sequent calculus $(\mathcal{L}(P_D),\vdash_D)$ can be defined  in the following three stages:

First, let  $$P_D=\set{\ua x}{x\in K^*(D)}$$ be the set of atomic formulae, $\mathrm{F}$ the syntactic constant for``false" and
 $$\mathcal{A}=\set{p_1,p_2,\ldots,p_n\vdash_D \mathrm{F}}{p_i\in P_D,p_1\cap p_2\cap \ldots\cap p_n=\emptyset}$$ the set of atomic disjointness assumptions.

 Second, define
  the set~$\mathcal{L}(P_D)$ of formulae by induction:
\begin{enumerate}
\item [{\em(L1)}] each atomic formula is an element of~$\mathcal{L}(P_D)$, and the constant connectives~$\mathrm{T}$ and $\mathrm{F}$ are elements of~$\mathcal{L}(P_D)$,
\item [{\em(L2)}] if $\varphi,\psi\in\mathcal{L}(P_D)$, then $\varphi\wedge\psi\in\mathcal{L}(P_D)$,
\item [{\em(L3)}] if a subset $\set{\varphi_i}{i\in I}$ of $\mathcal{L}(P_D)$ satisfies $\widehat{\varphi_i}\cap\widehat{\varphi_j}=\emptyset$ for all~$i\neq j\in I$, then $\dot{\bigvee}_{i\in I}\varphi_i\in \mathcal{L}(P_D)$,
    where $\widehat{\varphi}$ is the set generated by replacing the connectives $\mathrm{F},\mathrm{T}, \wedge$ and $\dot{\bigvee}$ in $\varphi$ by $\emptyset,D, \cup$ and $\dot{\bigcup}$, respectively.
\end{enumerate}

Finally, define the relation $\vdash_D$ by
\begin{equation}\label{equa3.3}
(\Gamma,\varphi)\in \vdash_D~\mathrm{if~and~only~if}~\widehat{\psi_1}\cap\widehat{\psi_2}\cap\cdots\cap\widehat{\psi_n}\subseteq \widehat{\varphi}
\end{equation}
   where, $\Gamma=\{\psi_1,\psi_2,\cdots,\psi_n\}\subseteq \mathcal{L}(P_D)$ and $\varphi \in\mathcal{L}(P_D)$.

\end{thm}
$\mathbf{Proof.}$

It is clear that $\widehat{p}=p$ for all $p\in P_D$, and $(\{p_1,p_2,\ldots,p_n\},\mathrm{F)\in \vdash_D}$ whenever $p_1,p_2,\ldots,p_n\vdash_D\mathrm{F}$ is an element of $\mathcal{A}$.

We first show that the class $\mathcal{L}(P_D)$ of all formulae  is closed under the rules of disjunctive formulae proposed in Definition~\ref{d2.3}. To this end, we claim that   $$\set{\widehat{\varphi}}{\varphi\in \mathcal{L}(P_D)}=\mathcal{U}(D).$$

 In fact, given a decomposable set $U\in \mathcal{U}(D)$,
  let

 ~~~~~~~~~~~~~~~~~~~~~ $\overleftarrow{U}=\left\{\begin{array}{ll}
\mathrm{T}, \ \ \ \ \ \ \ \ \ \ \ \ \ \ \ \ $if$~
U=D;\\
\mathrm{F},\ \ \  \ \ \ \ \ \ \ \ \ \ \ \ \ $if$~
U=\emptyset;\\
\dot{\bigvee}_{a\in A}\ua a, \ \ \ \ \ \ \  $if$~
U=\dot{\bigcup}_{a\in A}\ua a $~for~some~$A\subseteq K^*(D).\end{array}\right.$\\
Then by  the  construction of $\mathcal{L}(P_D)$, the expression $\overleftarrow{U}$ defined above is a member of  $\mathcal{L}(P_D)$.

 Assume that $\varphi\in \mathcal{L}(P_D)$, then by Definition \ref{dd3.6}, it is clear  that $\widehat{\varphi}\in \mathcal{U}(D)$, which means that $\set{\widehat{\varphi}}{\varphi\in \mathcal{L}(P_D)}\subseteq\mathcal{U}(D)$. For the converse inclusion, let $U\in \mathcal{U}(D)$. Then $\overleftarrow{U}\in \mathcal{L}(P_D)$, and hence $\widehat{\overleftarrow{U}}\in \set{\widehat{\varphi}}{\varphi\in \mathcal{L}(P_D)}$. This shows that $\mathcal{U}(D)\subseteq \set{\widehat{\varphi}}{\varphi\in \mathcal{L}(P_D)}$.

Now we  check the class $\mathcal{L}(P_D)$ of formulae is closed under the rules of disjunctive formulae proposed in Definition~\ref{d2.3}.

 From condition~(L1) and (L2), it follows that $\mathcal{L}(P_D)$ is closed under the rules (Const), (At) and (Conj). Since $D$ is an algebraic $L$-domain, the class $\mathcal{U}(D)$ is closed under disjoint union. This implies that $\mathcal{L}(P_D)$ is closed under the rule (Disj).

  Next, we have to check the rules of valid sequents. But all these rules can be shown routinely because of equation~(\ref{equa3.3}). We only illustrate this for the rule (Lwk):

  Assume that $\Gamma\vdash \phi$ is a valid sequent, where $\Gamma=\{\psi_1,\psi_2,\cdots,\psi_n\}$. Then $\widehat{\psi_1}\cap\widehat{\psi_2}\cap\cdots\cap\widehat{\psi_n}\subseteq \widehat{\phi}$. Thus $\widehat{\psi_1}\cap\widehat{\psi_2}\cap\cdots\cap\widehat{\psi_n}\cap \widehat{\psi}\subseteq \widehat{\phi}$ for all formula $\psi$. This means that the sequent $\Gamma,\psi\vdash \phi$ is valid.
 \hfill$\square$
\begin{prop}

 For the disjunctive sequent calculus $(\mathcal{L}(P_D),\vdash_D)$ associated with a given algebraic L-domain~$(D,\leq)$, we have
\begin{enumerate}
\item [{\em(1)}] a formula $\varphi$ is a tautology if and only if $\widehat{\varphi}=D$;
\item [{\em(2)}] a formula $\varphi$ is a contradiction if and only if $\widehat{\varphi}=\emptyset$;
\item [{\em(3)}] a formula $\varphi$ is a  satisfiable formula if and only if $\widehat{\varphi}=\ua A$, where $A$ is a nonempty pairwise inconsistent subset of $K^*(D)$.
\item [{\em(4)}] two formulae $\varphi$ and $\psi$ are logically equivalent if and only if $\widehat{\varphi}=\widehat{\psi}$:
\item [{\em(5)}] for any finite subset $\{ x_1, x_2,\cdots,x_n\}$ of $K^*(D)$, $\ua x_1\wedge \ua x_2\wedge\cdots\wedge \ua x_n$ is a conjunction  if and only if $\{x_1,x_2,\cdots,x_n\}$ has an upper bound in $D$.
\end{enumerate}
\end{prop}
$\mathbf{Proof.}$

 Straightforward from Theorem~\ref{p3.5}.
\hfill$\square$

For any nonempty proper subset $S$ of $\mathcal{L}(P_D)$, let
$$\widehat{S}=\set{\widehat{\varphi}}{\varphi\in S}.$$ Then we have
the following alternative
 characterization for a logical state of the disjunctive sequent calculi $(\mathcal{L}(P_D),\vdash_D)$.
\begin{prop}\label{p3.6}

Given an algebraic $L$-domain $(D,\leq)$, a subset
$S$ of $\mathcal{L}(P_D)$ is a logical state of $(\mathcal{L}(P_D),\vdash_D)$ if and only if there exists some $d\in D$ such that $\widehat{S}=\set{U\in \mathcal{U}(D)}{d\in U}$.
\end{prop}
$\mathbf{Proof.}$

  Let $S$ be a subset of $\mathcal{L}(P_D)$ satisfying $\widehat{S}=\set{U\in \mathcal{U}(D)}{d\in U}$ for some $d\in D$. We prove that $S$ is a logical state of $(\mathcal{L}(P_D),\vdash_D)$ by showing $S$  satisfies conditions (S1) and (S2).

   To prove condition (S1), assume that  $ \dot{\bigvee}_{i\in I}\mu_i$ is a flat formula in $S$. Since $d\in \dot{\bigcup}_{i\in I}\widehat{\mu_i}$, there exists some $i_0\in I$ such that $d\in\mu_{i_0}$. This implies that $\mu_{i_0}\in S$. For condition (S2), assume that  $\psi\in S[\vdash_D]$. By \Cref{eq3.1}, there exists some  finite subset $\Gamma=\{\varphi_1,\varphi_2,\cdots,\varphi_n\}$ of $S$ such that $\Gamma\vdash \psi$ is a valid sequent. If $\Gamma=\emptyset$, then $\psi\in$Tau$(P_D)$, and thus $d\in \widehat{\psi}=D$. If $\Gamma\neq\emptyset$, then $d\in\widehat{\varphi_1}\cap\widehat{\varphi_2}\cap\cdots\cap\widehat{\varphi_n}\subseteq \widehat{\psi}$. This implies that $d\in \widehat{\psi}$ and hence $\psi\in S$. Condition (S2) follows.

For the converse implication,
assume that $S$ is a logical state of $(\mathcal{L}(P_D),\vdash_D)$. Then $\widehat{S}\subseteq \mathcal{U}(D) $. We are now ready to look for an element $d_S$ of $D$ such that $\widehat{S}=\set{U\in \mathcal{U}(D)}{d_S\in U}$. 
 This process is divided into three steps.

 Step 1, for any  $\psi\in S$, noting that $\widehat{\psi}\neq \emptyset$, there exists some pairwise inconsistent subset $A$ of $K(D)$ such that $\widehat{\psi}=\dot{\bigcup}_{a\in A}\ua a$. By condition (S2), we have $\ua a_0\in S$ for some $a_0\in A$. This implies that $\bigcap \widehat{S}=\bigcap\set{\ua a}{\ua a\in S}$.

 Step 2, we prove  $\set{a}{\ua a\in S}$ is a directed set of $D$. Let $a_1,a_2\in \set{a}{\ua a\in S}$. Since $\ua a_1,\ua a_2\in S$, by part (1) of Proposition~\ref{p3.2},  $\ua a_1\cap\ua a_2 \in \widehat{S}$. This implies $\ua a_1\cap\ua a_2\neq\emptyset$. Suppose $\ua a_1\cap\ua a_2=\bigcup_{b\in B}\ua b$, where $B$ is a nonempty set of pairwise incompatible elements of $K(D)$. By condition~(S2), $\ua b\in S$ for some $b\in B$. Therefor, $b\in \set{a}{\ua a\in S}$ and $a_1,a_2\leq b$.

Step 3, put $d_S=\mathrm{sup}\set{a}{\ua a\in S}$. Then $d_S\in D$ and $\bigcap \widehat{S}=\bigcap\set{\ua a}{\ua a\in S}=\ua d_S$. Thus $d_S\in U$ for all $U\in \widehat{S}$. Therefore, $\widehat{S}\subseteq\set{U\in \mathcal{U}(D)}{d_S\in U}$. Conversely, for any $U\in \mathcal{U}(D)$ with $d_S\in U$, there exists some $d\in K(D)$ such that $d_S\in \ua d\subseteq U$. Since $d_S=\mathrm{sup}\set{a}{\ua a\in S}$ and $\set{a}{\ua a\in S}$ is directed,  $d\leq a_0$ for some $a_0\in \set{a}{\ua a\in S}$. Let $\varphi\in \mathcal{L}(P_D)$ such that $\widehat{\varphi}=U$. Then  $\ua a_0\subseteq \widehat{\varphi}$, and hence $\ua a_0\vdash \varphi$ is a valid sequent. By condition (S1), we have  $U\in \widehat{S}$.
\hfill$\square$

With the above preparations, we obtain the representation theorem of algebraic $L$-domains.

\begin{thm}\label{t3.2}

Each algebraic $L$-domain $(D,\leq)$  is isomorphic to $(|(\mathcal{L}(P_D),\vdash_D)|,\subseteq)$.
\end{thm}
$\mathbf{Proof.}$

Define a function as follows:
 $$f:(D,\leq)\rightarrow (|(\mathcal{L}(P_D),\vdash_D)|,\subseteq),$$
 $$d\mapsto\set{\varphi\in \mathcal{L}(P_D)}{d\in \widehat{\varphi}}.
$$
 By Proposition~\ref{p3.6}, the set $\set{\varphi\in \mathcal{L}(P_D)}{d\in \widehat{\varphi}}$ is a logical state of $(\mathcal{L}(P_D),\vdash_D)$, then the function $f$ is  well-defined. For all logical state $S$ of $(\mathcal{L}(P_D),\vdash_D)$, as we have known in the proof of Proposition~\ref{p3.6}, it is not difficult to see that $f(d_S)=S$, where $d_S=\mathrm{sup}\set{a}{\ua a\in S}$.  This means that the function $f$ is surjective.

Moreover, it is easy to show that
 $d_1\leq d_2$ if and only if $$ \set{\varphi\in \mathcal{L}(P_D)}{\widehat{\varphi}\in \mathcal{U}(D), d_1\in \widehat{\varphi}}\subseteq \set{\varphi\in \mathcal{L}(P_D)}{\widehat{\varphi}\in \mathcal{U}(D), d_2\in \widehat{\varphi}}.$$ Therefore, the function $f$ is an order-isomorphism from $(D,\leq)$ to $(|(\mathcal{L}(P_D),\vdash_D)|,\subseteq)$. 
 \hfill$\square$

  We now see that there are technical advantages to investigating disjunctive sequent calculi rather than its Lindenbaum algebra. First, logical states of a disjunctive sequent calculus use the language of set theory. Second,
the features of algebraic $L$-domains can be  obtained directly by logical inference.  It further demonstrates its advantages by representing the category of algebraic $L$-domains with  Scott-continuous functions as morphisms in the next section.

 \section{A Categorical view}\label{4}

From a categorical viewpoint,  \Cref{3.2} has built an object part correspondence between algebraic $L$-domains and disjunctive sequent calculi.
  We now aim to look for  appropriate morphisms so as to obtain an categorical equivalence.

\subsection{Expressive disjunctive sequent calculi}
In the rest of the article, a special class of disjunctive sequent calculi  are considered as objects to construct a category equivalent to $\mathbf{ALD}$.

As we have always seen  in Proposition \ref{p3.5},  each algebraic $L$-domain~$(D,\leq)$  associated with a  disjunctive sequent calculus $(\mathcal{L}(P_D),\vdash_D)$. In this case,  a conjunction is of the form  $\ua x_1\wedge \ua x_2\wedge\cdots\wedge \ua x_n$, where $\{ x_1, x_2,\cdots,x_n\}\subseteq K^*(D)$ and $\{x_1,x_2,\cdots,x_n\}$ has an upper bound in $D$.

The following result gives a sufficient and necessary condition for  a conjunction  being irreducible in  $(\mathcal{L}(P_D),\vdash_D)$.

\begin{prop} \label{p3.8}

Let $(D,\leq)$ be an algebraic $L$-domain and $\{x_1,x_2,\cdots,x_n\}$ a nonempty subset of $K^*(D)$.  Then a conjunction $\ua x_1\wedge\ua x_2\wedge\cdots\wedge \ua x_n$ in  $(\mathcal{L}(P_D),\vdash_D)$ is irreducible if and only if $\mathrm{sup}\{x_1,x_2,\cdots,x_n\}\in D$.
\end{prop}
$\mathbf{Proof.}$

 For any conjunction $\ua x_1\wedge\ua x_2\wedge\cdots\wedge \ua x_n$,  we have seen that the set $\{x_1,x_2,\cdots,x_n\}$ has an upper bound in $D$. Since $D$ is an algebraic $L$-domain, there is a nonempty pairwise inconsistent subset $A$ of $K^*(D)$ such that $\ua x_1\wedge\ua x_2\wedge\cdots\wedge \ua x_n$ is logically equivalent to $\dot{\bigvee}_{a\in A}\ua a$ in $(\mathcal{L}(P_D),\vdash_D)$.
In this case, the set $A$ just is the minimal upper bounds of $\{ x_1, x_2,\cdots, x_n\}$.

Assume that $\mathrm{sup}\{x_1,x_2,\cdots,x_n\}\in D$, then $A$ must be a singleton, say $\{d\}$. The elements $x_1,x_2,\cdots,x_n$ are compact in $D$ and so is  $d$. This implies that $\ua d$ is a formula in  $(\mathcal{L}(P_D),\vdash_D)$, and thus, $\ua x_1\wedge\ua x_2\wedge\cdots\wedge \ua x_n$ is logically equivalent to $\ua d$ in  $(\mathcal{L}(P_D),\vdash_D)$. Let $\ua x_1\wedge\ua x_2\wedge\cdots\wedge \ua x_n\vdash\dot{\bigvee}_{i\in I}\phi_{i}$ be a valid sequent, where $\widehat{\phi_{i}}\cap\widehat{\phi_{j}}=\emptyset$ for all $i\neq j\in I$. Then
 $\ua d\subseteq \dot{\bigcup}_{i\in I}\widehat{\phi_{i}}$, and hence there exists  some $i_0\in I$ such that $\ua d\subseteq\widehat{\phi_{i_0}}$. Therefore, $\ua x_1\wedge\ua x_2\wedge\cdots\wedge \ua x_n\vdash\phi_{i_0}$ is valid. So $\ua x_1\wedge\ua x_2\wedge\cdots\wedge \ua x_n$ is an irreducible conjunction, as required.

  Conversely, if $\mathrm{sup}\{x_1,x_2,\cdots,x_n\}$ does not exists in $K^*(D)$, then  there are at least two elements in $A$. Therefore, $\bigcup_{a\in A}\ua a$ is not a subset of $\ua a$ for all $a\in A$. Thus $\dot{\bigvee}_{a\in A}\ua a \vdash \ua a$ is not a valid sequent. But $\dot{\bigvee}_{a\in A}\ua a \vdash \dot{\bigvee}_{a\in A}\ua a$ is valid. Hence $\ua x_1\wedge\ua x_2\wedge\cdots\wedge \ua x_n$ is not an irreducible conjunction.
\hfill$\square$.

 \begin{defn}\label{d3.6}

A disjunctive sequent calculus $(\mathcal{L}(P),\vdash)$ is said to be expressive if, for any  satisfiable formula $\psi$, there exists an irreducible flat formula $\dot{\bigvee}_{i\in I}\mu_{i}$  such that $\psi\vdash \dot{\bigvee}_{i\in I}\mu_{i}$ is valid and $\mu_{i}\vdash\psi$ are  valid for all $i\in I$.
\end{defn}

\begin{prop}\label{p3.9}

For any algebraic $L$-domain~$(D,\leq)$, the associated disjunctive sequent calculus $(\mathcal{L}(P_D),\vdash_D)$ is expressive.
\end{prop}
$\mathbf{Proof.}$

 Let  $(\mathcal{L}(P_D),\vdash_D)$ be the disjunctive sequent calculus associated with an algebraic $L$-domian~$(D,\leq)$. Proposition~\ref{p3.8} reveals that every atomic formula of  $(\mathcal{L}(P_D),\vdash_D)$ is an irreducible conjunction.
For any  satisfiable formula $\psi$ of  $(\mathcal{L}(P_D),\vdash_D)$, the set $\widehat{\psi}$ has a form $\ua A$, where $A$  is a nonempty pairwise inconsistent subset  of $K^*(D)$. Then  $\dot{\bigvee}_{a\in A}\ua a$ is a  flat formula such that $\psi\vdash\dot{\bigvee}_{a\in A}\ua a$ and $\ua a\vdash \psi$ are valid for all $a\in A$.
\hfill$\square$

With respect to an expressive  disjunctive sequent calculus~ $(\mathcal{L}(P),\vdash)$, a logical state has another alternative characterization.

\begin{thm}\label{p3.10}

 Let $(\mathcal{L}(P),\vdash)$ be an expressive  disjunctive sequent calculus and $S$ a nonempty proper subset of $\mathcal{L}(P)$.
 Then $S$ is a logical state if and only if  the set $\set{\{\mu\}[\vdash]}{\mu\in S\cap\mathcal{IC}(P)}$ is directed  and $S$ is its union.
\end{thm}
$\mathbf{Proof.}$

  Assume that  the set  $\set{\{\mu\}[\vdash]}{\mu\in S\cap\mathcal{IC}(P)}$ is directed  and $S$ is its union. We may appeal to  part~(2) of Proposition~\ref{p3.3} and Proposition~\ref{p3.7} to deduce that $S$ is a logical state.
So that the only interesting thing is the ``only if '' part of the proof.

  Let $S$ be a logical  state.
  We first claim that for any $\Gamma\sqsubseteq S$ there exists some  $\mu\in S\cap\mathcal{IC}(P)$ such that $[\Gamma]_S\subseteq\{\mu\}[\vdash]$. Indeed, if $\Gamma=\emptyset$, then $[\Gamma]_S\subseteq\{\mu\}[\vdash]$ for any $\mu\in S\cap\mathcal{IC}(P)$. If $\Gamma\neq\emptyset$, then $\Gamma\sqsubseteq S$ implies that $\bigwedge\Gamma\in S$. Since $(\mathcal{L}(P),\vdash)$ is an expressive  disjunctive sequent calculus, there exists some irreducible flat formula $\dot{\bigvee}_{i\in I}\mu_{i}$ such that $\bigwedge\Gamma\vdash\dot{\bigvee}_{i\in I}\mu_{i}$ and $\mu_i\vdash\bigwedge\Gamma$ are  valid for all $i\neq j\in I$. Then $\dot{\bigvee}_{i\in I}\mu_{i}\in S$, and hence $\mu_{i_0}\in S$ for some $i_{0}\in I$. From $\mu_{i_0}\vdash \bigwedge\Gamma$ it follows that $\Gamma\subseteq \{\mu_{i_0}\}[\vdash]$, which implies that $[\Gamma]_S\subseteq\{\mu\}[\vdash]$.

We next claim that for any $\mu\in S\cap\mathcal{IC}(P)$ there exists some $\Gamma\subseteq S$ such that $\{\mu\}[\vdash]\subseteq [\Gamma]_S$.
Indeed, taking $\Gamma=\{\mu\}$,  since $[\Gamma]_S$ is a logical state, $\{\mu\}[\vdash]\subseteq[\Gamma]_S$, by condition (S1).

Finally, as we have  seen from the proof of Theorem~\ref{t3.1},  the set $\set{[\Gamma]_S}{ \Gamma\sqsubseteq S}$ is directed and $S$ is its union. Then  the set $\set{\{\mu\}[\vdash]}{ \mu\in S\cap\mathcal{IC}(P)}$ is directed and $S$ is its union.
\hfill$\square$

\subsection{Morphisms between expressive disjunctive sequent calculi}
 A main contribution of this subsection is the introduction of a notion of consequence relation between expressive disjunctive sequent calculi, which can be used to represent Scott-continuous functions between algebraic $L$-domains.

 \begin{defn}\label{d4.1}

  Let $(\mathcal{L}(P),\vdash_P)$ and $(\mathcal{L}(P),\vdash_Q)$ be two expressive disjunctive sequent calculi. 
  A binary relation  $\Theta\subseteq \mathcal{IC}(P)\times\mathcal{L}({Q})$ is called a consequence relation from $(\mathcal{L}(P),\vdash_P)$ to $(\mathcal{L}(P),\vdash_Q)$ if, for all $\mu\in \mathcal{IC}(P)$ and $\psi \in \mathcal{L}({Q})$, the following conditions hold:
\begin{enumerate}
\item [{\bf(R1)}] whenever there is some $(\nu,\psi)\in\Theta$ such that  $\mu\vdash_P \nu$ is a valid sequent, then  $(\mu,\psi)\in\Theta$;
\item [{\bf(R2)}]  whenever there is some $(\mu,\varphi)\in\Theta$  such that  $\varphi\vdash_Q\psi$ is a valid sequent, then  $(\mu,\psi)\in\Theta$;

\item [{\bf(R3)}] whenever $(\mu,\psi)\in \Theta$, then $(\mu,\nu)\in \Theta$ and the sequent  $\nu\vdash_Q\psi$ is valid for some  $\nu\in \mathcal{IC}({Q})$.
\end{enumerate}
\end{defn}
We use $\Theta:(\mathcal{L}(P),\vdash_P)\rightarrow(\mathcal{L}(P),\vdash_P)$ to denote the consequence relation $\Theta$ from $(\mathcal{L}(P),\vdash_P)$ to $(\mathcal{L}(P),\vdash_Q)$ defined above.
\begin{prop} \label{r4.1}

Let $\Theta:(\mathcal{L}(P),\vdash_P)\rightarrow(\mathcal{L}(P),\vdash_P)$ be a consequence relation.  Then  the following conditions are  equivalent.
\begin{enumerate}
\item [{\em(1)}] $(\mu,\psi)\in\Theta$.
\item [{\em(2)}] There exists some $\nu\in \mathcal{IC}({P})$  such that $\mu\vdash_P \nu$ is valid and $(\nu,\psi)\in\Theta$.
\item [{\em(3)}]  There exists some  $\varphi\in \mathcal{L}({Q})$ such that $\varphi\vdash_Q\psi$ is valid and $(\mu,\varphi)\in\Theta$.
\item [{\em(4)}] There exists $\nu\in \mathcal{IC}({Q})$ such that $\nu\vdash_Q\psi$ is valid and $(\mu,\nu)\in \Theta$.
\end{enumerate}
\end{prop}
$\mathbf{Proof.}$

 Straightforward from Definition~\emph{\ref{d4.1}}.
\hfill$\square$

For all consequence relation  $\Theta$ from $(\mathcal{L}(P),\vdash_P)$ to $(\mathcal{L}(P),\vdash_P)$
  and all subset $X$ of $\mathcal{L}({P})$, set
\begin{equation}\label{eq4.1}
 \Theta[X]=\set{\varphi \in \mathcal{L}({Q})}{ (\exists  \mu\in X\cap \mathcal{IC}({P}) ) (\mu,\varphi)\in\Theta}.
\end{equation}
Then it is clear that $\Theta[X_1] \subseteq\Theta[X_2]$ for any  $X_1\subseteq X_2\subseteq \mathcal{L}({P})$.

The following proposition shows that a consequence relation provides a passage from logical states of an expressive disjunctive sequent calculus to those of another one.
\begin{prop} \label{p4.1}

Let $\Theta$ be a consequence relation from $(\mathcal{L}(P),\vdash_P)$ to $(\mathcal{L}(P),\vdash_P)$. 
\begin{enumerate}
\item [\emph{(1)}] If $S$ is a logical state of $(\mathcal{L}(P),\vdash_P)$,
then $\Theta[S] $ is a logical state of $(\mathcal{L}(Q),\vdash_Q)$.
\item [\emph{(2)}] If $\mu\in\mathcal{IC}({P})$, then $\Theta[\{\mu\}]=\Theta[\{\mu\}[\vdash_P]]$.
\end{enumerate}
\end{prop}
$\mathbf{Proof.}$

(1)
  Let $\dot{\bigvee}_{i\in I}\mu_i\in \Theta[S]\cap\mathcal{N}({P})$. Then there exists $\mu\in S\cap \mathcal{IC}({P})$ such that $(\mu,\dot{\bigvee}_{i\in I}\mu_i)\in\Theta$. By condition~(R3), there exists some $\nu\in \mathcal{IC}({Q})$ such that $(\mu,\nu)\in \Theta$ and $\nu\vdash_Q\dot{\bigvee}_{i\in I}\mu_i$ is valid. Since $\nu$ is an irreducible conjunction,  $\nu\vdash_Q\mu_{i_0}$ is valid for some $i_0\in I$. Using condition~(R2) again,  $(\mu,\mu_{i_0})\in \Theta$, and thus $\mu_{i_0}\in \Theta[S]$.

 Assume that $\varphi\in (\Theta[S])[\vdash_Q]$. Then there exist $\mu\in S\cap \mathcal{IC}({P})$ and $\psi\in \mathcal{L}({Q})$ such that $(\mu,\psi)\in\Theta$ and $\psi\vdash_Q\varphi$ is valid. By condition~(R2), we have $(\mu,\varphi)\in\Theta$. This implies that $\varphi\in \Theta[S]$ and hence $(\Theta[S])[\vdash_Q]\subseteq \Theta[S]$.

(2) It is clear $\Theta[\{\mu\}]\subseteq\Theta[\{\mu\}[\vdash_P]]$. Conversely, let $\varphi\in \Theta[\{\mu\}[\vdash_P]]$. Then there exists some $\nu\in \{\mu\}[\vdash_P]\cap \mathcal{IC}({P})$ such that $(\nu,\varphi)\in \Theta$. But $\nu\in \{\mu\}[\vdash_P]$ implies that $\mu\vdash_P\nu$ is valid. By condition~(R1), we have $(\mu,\varphi)\in \Theta$, and thus $\varphi\in \Theta[\{\mu\}]$.
\hfill$\square$

Now we consider the relationship between consequence relations and
 Scott-continuous functions.
\begin{thm} \label{t4.1}

Let $(\mathcal{L}(P),\vdash_P)$ and $(\mathcal{L}(Q),\vdash_Q)$ be expressive disjunctive sequent calculi.
 For all consequence relation $\Theta: (\mathcal{L}(P),\vdash_P)\rightarrow(\mathcal{L}(Q),\vdash_Q)$, define a function
$f_{\Theta}:|(\mathcal{L}(P),\vdash_P)|\rightarrow |(\mathcal{L}(Q),\vdash_Q)|$ by
\begin{equation}\label{e4.2}
f_{\Theta}(S)=\Theta[S].
\end{equation}
Then $f_{\Theta}$ is Scott-continuous.

 Conversely, for all Scott-continuous function $f : |(\mathcal{L}(P),\vdash_P)|\rightarrow |(\mathcal{L}(Q),\vdash_Q)|$, define
$\Theta_{f}\subseteq \mathcal{C}_0({P})\times\mathcal{L}({Q})$ by
 \begin{equation}\label{e4.3}
(\mu,\psi)\in\Theta_f\Leftrightarrow \psi\in f(\{\mu\}[\vdash_P]).
\end{equation}
Then $\Theta_{f}$ is a consequence relation from $ (\mathcal{L}(P),\vdash_P)$ to $(\mathcal{L}(Q),\vdash_Q)$.
Moreover, $\Theta_{f_{\Theta}}=\Theta$ and  $f_{\Theta_{f}}=f$.
\end{thm}

$\mathbf{Proof.}$

 We divided the proof  into three claims.

Claim 1: The function $f_{\Theta}$ is Scott-continuous.

 By part (2) of Proposition \ref{p4.1}, the function $f_{\Theta}$ is well-defined. Let $\set{S_i}{i\in I}$ be a directed subset of logical states. Then $\bigcup_{i\in I}S_i$ is a logical state. From \Cref{eq3.1} it follows that the function $f_{\Theta}$ is monotone, and hence $\bigcup_{i\in I}f_{\Theta}(S_i)\subseteq f_{\Theta}(\bigcup_{i\in I}S_i)$.  To prove the function $f_{\Theta}$ is Scott-continuous, it suffices to show that $ f_{\Theta}(\bigcup_{i\in I}S_i)\subseteq \bigcup_{i\in I}f_{\Theta}(S_i)$. If $\varphi\in f_{\Theta}(\bigcup_{i\in I}S_i)=\Theta[\bigcup_{i\in I}S_i]$, then there exists some $\mu\in \bigcup_{i\in I}S_i\cap \mathcal{IC}(P)$ such that $(\mu,\varphi)\in \Theta$. From $\mu\in\bigcup_{i\in I}S_i$, it follows that $\mu\in S_{i_0}$ for some $i_0\in I$. Thus $\varphi\in f_{\Theta}(S_{i_0})$, and therefore, $ f_{\Theta}(\bigcup_{i\in I}S_i)\subseteq \bigcup_{i\in I}f_{\Theta}(S_i)$.

Claim 2. $\Theta_{f}$ is a consequence relation from $ (\mathcal{L}(P),\vdash_P)$ to $(\mathcal{L}(Q),\vdash_Q)$.

 It suffices to show that $\Theta_f$ satisfies conditions~(R1--R3).

For condition (R1),
assume that  $\nu\in \mathcal{IC}(P)$  such that $\mu\vdash_P \nu$ is valid and $(\nu,\psi)\in\Theta_f$. Then $\psi\in f(\{\nu\}[\vdash_P])$. From $\mu\vdash_P \nu$ it follows that $\{\nu\}[\vdash_P]\subseteq \{\mu\}[\vdash_P]$. Since $f$ is monotone, $\psi\in f(\{\mu\}[\vdash_P])$. This means that $(\mu,\psi)\in\Theta_f$.

For condition (R2),
assume that $(\mu,\varphi)\in\Theta_f$  and $\varphi\vdash_Q\psi$ is valid. Then
$\varphi\in f(\{\mu\}[\vdash_P])$. Since $f(\{\mu\}[\vdash_P])$ is a logical state and $\varphi\vdash_Q\psi$ is valid, $\psi\in f(\{\mu\}[\vdash_P])$. That is $(\mu,\psi)\in\Theta_f$.

 For condition (R3),
assume that $(\mu,\psi)\in \Theta_f$. Then $\psi\in f(\{\mu\}[\vdash_P])$. Since the disjunctive sequent calculus $(\mathcal{L}(Q),\vdash_Q)$ is expressive, there exists an irreducible flat formula $\dot{\bigvee}_{i\in I}\mu_{i}$  such that $\psi\vdash \dot{\bigvee}_{i\in I}\mu_{i}$ and $\mu_{i}\vdash\psi$ are all valid for all $i\in I$. Note that $f(\{\mu\}[\vdash_P])$ is a logical state, it follows that $\dot{\bigvee}_{i\in I}\mu_{i}\in f(\{\mu\}[\vdash_P]) $. Thus $\mu_{i_0}\in f(\{\mu\}[\vdash_P])$ for some $i_0\in I$. Let $\mu_{i_0}=\nu$.
  Then we obtain some $\nu\in \mathcal{IC}(Q)$ such that $(\mu,\nu)\in \Theta$ and $\nu\vdash_Q\psi$ is valid.

Claim 3. $\Theta_{f_{\Theta}}=\Theta$ and  $f_{\Theta_{f}}=f$.

 For any $\mu\in\mathcal{IC}(P)$ and $\varphi\in \mathcal{L}(Q)$, we have
\begin{align*}
(\mu,\varphi)\in\Theta_{f_{\Theta}}&\Leftrightarrow \varphi\in f_{\Theta}(\{\mu\} [\vdash_P])\\
&\Leftrightarrow \varphi\in \Theta[\{\mu\} [\vdash_P]]\\
&\Leftrightarrow (\exists\nu\in\mathcal{IC}(Q)((\mu,\varphi)\in\Theta,
\mu\vdash_P\nu\in{\bf T}(P)))\\
&\Leftrightarrow (\mu,\varphi)\in\Theta.
\end{align*}
This proves that $\Theta_{f_{\Theta}}=\Theta$.

  For any $S \in |(\mathcal{L}(P),\vdash_P)|$, we have
\begin{align*}
f_{\Theta_{f}}(S)&=\Theta_{f}[S]\\
&=\set{\varphi\in \mathcal{L}(Q)}{(\exists \mu\in S\cap\mathcal{IC}(P))(\mu,\varphi)\in\Theta_{f}}\\
&=\set{\varphi\in \mathcal{L}(Q)}{(\exists \mu\in S\cap\mathcal{IC}(P))\varphi\in f(\{\mu\} [\vdash_P])}\\
&=\bigcup \set{f(\{\mu\} [\vdash_P])}{\mu\in S\cap\mathcal{IC}(P)}\\
&=f(\bigcup\set{\{\mu\} [\vdash_P]}{\mu\in S\cap\mathcal{IC}(P)}\\
&=f(S).
\end{align*}
This proves that $f_{\Theta_{f}}=f$.
\hfill$\square$

Similar to the result presented in Theorem~\ref{t4.1},  there is also a one-to-one correspondence between Scott-continuous functions from algebraic $L$-domain $(D_1,\leq_1)$ to algebraic $L$-domain $(D_2,\leq_2)$ and  consequence relations from $(\mathcal{L}(P_{D_1}),\vdash_{D_1})$ to $(\mathcal{L}(P_{D_2}),\vdash_{D_2})$. But familiarity with this is not essential for our investigation that follows, so we
omit the proof and only state the result.

\begin{thm}

Let $(D_1,\leq_1)$ and $(D_2,\leq_2)$ be algebraic $L$-domains.
 For all Scott-continuous function $h: D_1\rightarrow D_2$, define
$\Omega_{h}\subseteq \mathcal{IC}(P_{D_1})\times\mathcal{L}(P_{D_2})$ by
\begin{equation}
(\mu,\psi)\in\Omega_h\Leftrightarrow  f(\mu)\subseteq\psi.
\end{equation}

Then $\Omega_{h}$ is a consequence relation from $(\mathcal{L}(P_{D_1}),\vdash_{D_1})$ to $ (\mathcal{L}(P_{D_2}),\vdash_{D_2})$.

 Conversely, for any consequence relation $\Omega : (\mathcal{L}(P_{D_1}),\vdash_{D_1})\rightarrow (\mathcal{L}(P_{D_2}),\vdash_{D_2})$, define an assignment~$h_{\Omega}$ by
\begin{equation}
\begin{aligned}
h_{\Omega}(x)=&\mathrm{sup}\set{y\in K(D_2)}{\ua y\in \set{U}{(\exists \mu\in\mathcal{IC}(P_{D_1}))( x\in\mu,(\mu,U)\in \Omega)}}.
\end{aligned}
\end{equation}
Then $h_{\Omega}$ is a Scott-continuous function from $ D_1$ to $D_2$.
Moreover, $\Omega_{h_{\Omega}}=\Omega$ and  $h_{\Omega_{h}}=h$.
\end{thm}

\subsection{Categorical equivalence}

It remains to consider the category of expressive disjunctive sequent calculi. What we want to do is to show this category is equivalent to the category $\mathbf{ALD}$ of algebraic $L$-domain with Scott continuous functions as morphisms.
\begin{prop} \label{p4.2}

Expressive disjunctive sequent  calculi with consequence relations form a category $\mathbf {EDSC}$.
\end{prop}
$\mathbf{Proof.}$

Let $\Theta$ be a consequence relation from $(\mathcal{L}(P),\vdash_P)$ to  $(\mathcal{L}(Q),\vdash_Q)$, and $\Theta'$  a consequence relation from $(\mathcal{L}(Q),\vdash_Q)$ to  $(\mathcal{L}(R),\vdash_R)$. Define  $\Theta'\circ \Theta\subseteq \mathcal{IC}(P)\times\mathcal{L}(R)$ by
\begin{equation}\label{e4.8}
(\mu,\varphi)\in\Theta'\circ \Theta\Leftrightarrow
(\exists \nu\in\mathcal{IC}(Q))((\mu,\nu)\in\Theta,(\nu,\varphi)\in\Theta'),
\end{equation}
and id$_P\subseteq \mathcal{IC}(P)\times\mathcal{L}(P)$ by
\begin{equation}\label{e4.9}
(\mu,\varphi)\in\text{id}_P \Leftrightarrow \varphi\in \{\mu\}[\vdash_P].
\end{equation}
Then routine checks verify that $\Theta'\circ \Theta$ is a consequence relation  from $(\mathcal{L}(P),\vdash_P)$ to $(\mathcal{L}(R),\vdash_R)$ and
 $\text{id}_P$ is a consequence relation from $(\mathcal{L}(P),\vdash_P)$ to itself.

  Using the same argument as checking the associative law of a traditional relation composition, we can carry out the composition~$\circ$ defined by~\Cref{e4.8} is  associative. Conditions (R1) and (R2)  yield that $\text{id}_P$ is the identity morphism of $(\mathcal{L}(P),\vdash_P)$. So
  $\mathbf{EDSC}$ is a category, as required.
\hfill$\square$

We use the following well known fact to establish a categorical equivalence.

\begin{lemma}\label{l4.1} (\cite{b27})

Let $\mathbf{C}$ and $\mathbf{D}$ be two categories. Then $\mathbf{C}$ and $\mathbf{D}$ are categorically equivalent if and only if there exists a functor~$\mathcal{G}:\mathbf{C}\rightarrow\mathbf{D}$ such that $\mathcal{G}$ is full, faithful and essentially surjective on objects, that is for every object~$D$ of $\mathbf{D}$, there exists some object $C$ of $\mathbf{C}$ such that $\mathfrak{F}(C)\cong D$.
\end{lemma}

\begin{prop}\label{p4.3}

$\mathcal{G}:\mathbf{EDSC}\rightarrow \mathbf{ALD}$ is a functor which maps every expressive disjunctive sequent calculi $(\mathcal{L}(P),\vdash_P)$ to $(|(\mathcal{L}(P),\vdash_P)|,\subseteq)$ and consequence relation $\Theta: (\mathcal{L}(P),\vdash_P)\rightarrow (\mathcal{L}(Q),\vdash_Q)$ to $f_{\Theta}$, where $f_{\Theta}$ is defined by equation~\emph{(\ref{e4.2})}.
\end{prop}
$\mathbf{Proof.}$

By Theorems~\ref{t3.1} and \ref{t4.1}, $\mathcal{G}$ is well-defined.
For all $S\in|(\mathcal{L}(P),\vdash_P)|$, we have
\begin{align*}
\mathcal{G}(\text{id}_{P})(S)&=f_{\text{id}_P}(S)\\
&=\text{id}_P[S]\\
&=\set{\varphi\in \mathcal{L}(P)}{(\exists \mu\in S\cap\mathcal{IC}(P))(\mu,\varphi)\in \text{id}_P)}\\
&=\set{\varphi\in \mathcal{L}(P)}{(\exists \mu\in S\cap\mathcal{IC}(P))\varphi\in \{\mu\}[\vdash_P]}\\
&=\bigcup\set{\{\mu\}[\vdash_P]}{\mu\in S\cap\mathcal{IC}(P)}\\
&=S.
\end{align*}
This implies that $\mathcal{G}$ preserves the identity morphism.

Let $\Theta: (\mathcal{L}(P),\vdash_P)\rightarrow (\mathcal{L}(Q),\vdash_Q)$ and $ \Theta':(\mathcal{L}(Q),\vdash_Q)\rightarrow (\mathcal{L}(R),\vdash_R)$ be  consequence relations.
For all $S\in|(\mathcal{L}(P),\vdash)|$, we have
\begin{align*}
\mathcal{G}(\Theta'\circ\Theta)(S)&= f_{\Theta'\circ\Theta}(S) \\
&=\Theta'\circ\Theta[S] \\
&=\set{\varphi}{(\exists \mu\in S\cap\mathcal{IC}(P))(\mu,\varphi)\in \Theta'\circ\Theta}\\
&=\set{\varphi}{(\exists \mu\in S\cap\mathcal{IC}({P}), \exists \nu\in\mathcal{IC}({Q}))((\mu,\nu)\in \Theta,(\nu,\varphi)\in \Theta')}\\
&=\set{\varphi}{( \exists \nu\in\mathcal{IC}({Q}))(\nu\in f_{\Theta}(S),(\nu,\varphi)\in \Theta')}\\
&=\set{\varphi}{( \exists \nu\in\mathcal{IC}({Q})\cap f_{\Theta}(S))(\nu,\varphi)\in \Theta'}\\
&=f_{\Theta'}(f_{\Theta}(S)).
\end{align*}
This implies that $\mathcal{G}(\Theta'\circ\Theta)=\mathcal{G}(\Theta')\circ\mathcal{G}(\Theta)$, and then $\mathcal{G}$ preserves the composition.
\hfill$\square$

\begin{thm}\label{t4.3}

$\mathbf{EDSC}$ and $\mathbf{ALD}$ are categorically equivalent.
\end{thm}
$\mathbf{Proof.}$

According to Theorem~\ref{t3.2} and Lemma~\ref{l4.1}, it suffices to show that the functor~$\mathcal{G}$ defined in Proposition~\ref{p4.3} is full and faithful.

 For any Scott-continuous function $f : |(\mathcal{L}(P),\vdash_P)|\rightarrow |(\mathcal{L}(Q),\vdash_Q)|$,  by Theorem~\ref{t4.1}, the relation $\Theta_{f}$ defined by equation~(\ref{e4.3})
is a consequence relation from $ (\mathcal{L}(P),\vdash_P)$ to $(\mathcal{L}(Q),\vdash_Q)$ and $\mathcal{G}(\Theta_f)=f_{\Theta_{f}}=f$.
This implies that $\mathcal{G}$ is full.

 Let $\Theta_1,\Theta_2: (\mathcal{L}(P),\vdash_P)\rightarrow (\mathcal{L}(Q),\vdash_Q)$ be two consequence relations with $f_{\Theta_1}=f_{\Theta_2}$, where $f_{\Theta_1}$ and $f_{\Theta_2}$ are defined by equation~(\ref{e4.2}). For any $\mu\in \mathcal{IC}(P)$, since
\begin{align*}
(\mu,\varphi)\in\Theta_1&\Leftrightarrow \varphi\in\Theta_1[\{\mu\}]\\
&\Leftrightarrow \varphi\in\Theta_1[\{\mu\}[\vdash_P]]\\
&\Leftrightarrow \varphi\in f_{\Theta_1}(\{\mu\}[\vdash_P])\\
&\Leftrightarrow \varphi\in f_{\Theta_2}(\{\mu\}[\vdash_P])\\
&\Leftrightarrow (\mu,\varphi)\in\Theta_2,
\end{align*}
it follows that $\Theta_1=\Theta_2$, and hence $\mathcal{G}$ is faithful.
\hfill$\square$

Combining Theorem~\ref{t4.3}  with the fact that $\mathbf {ALD}$ is a cartesian closed category, we have the following result:
\begin{cor}

$\mathbf {EDSC}$ is a cartesian closed category.
\end{cor}

\section{Funding} This study was funded by the National Natural Science Foundation of China(11771134).


\begin{thebibliography}{100}

\bibitem{b1} D. S. Scott, Outline of a mathematical theory of Computation, in: Proc.
4th Annual Princeton Conference on Information Sciences and Systems, Princeton, USA, 1970.

\bibitem{b2} D. S. Scott, C. Strachey, Towards a mathematical semantics for computer languages, in:  21st Symposium on Computers and Automata, Brooklyn, USA, 1971.

\bibitem{b3} M. Ern$\acute{\text{e}}$, Categories of locally hypercompact spaces and quasicontinuous posets, Applied Categorical Structures,  26 (2018) 823--854.

\bibitem{b4} P.~Hitzler, M.~Kr$\ddot{o}$etzsch, G.~Zhang, A categorical view on algebraic lattices in formal concept analysis, Fundamenta Informaticae,  74 (2004) 1--29.

\bibitem{b5} W.~Ho, J.~Goubault-Larrecq, A.~Jung, X.~Xi, The Ho-Zhao problem, Logical Methods in Computer Science, 14 (2016) 1--19.

\bibitem{b6} W. Yao, A categorical isomorphism between injective stratified fuzzy $T_0$ spaces and fuzzy continuous lattices, LIEEE Transactions on Fuzzy Systems,  24 (2016) 131--139.


\bibitem{b7} J. Lu, B. Zhao, K. Wang, SI-continuous spaces and continuous posets, Topology and its Applications, 264 (2019) 313--321.

\bibitem{b8}  K. Keimel, J. Lawson, D-completions and the d-topology, Annals of Pure and Applied Logic, 159 (2009)  292--306.

\bibitem{b9} D.~S. Scott, Domains for denotational semantics, Lecture Notes in Computer Science, 140 (1982) 577--613.

\bibitem{b10} S.~Abramsky, Domain theory and the logic of observable properties, Ph.D. dissertation,   Univ. of London, London,  UK, 1987.


\bibitem{b11} S.~Abramsky,  Domain theory in logical form, Annals of Pure and Applied Logic, 51 (1991) 1--77.

\bibitem{b12} K.~G. Larsen, G.~Winskel, Using information systems to solve recursive domain equations
  effectively, Lecture Notes in Computer Science, 173, (1984) 109--130.

\bibitem{b13} M.~Huang, X.~Zhou, Q.~Li, Re-visiting axioms of information systems,  Information and Compution, 247 (2015) 130--140.

\bibitem{b14} D.~Spreen, L.~Xu, X.~Mao, Information systems revisited: the general continuous case, Theoretical Computer Science, 405 (2008) 176--187.

\bibitem{b15} Q. He, L.~Xu, Weak algebraic information systems and a new equivalent category of DOM of domains, Theoretical Computer Science, 763 (2019) 1--11.

\bibitem{b16} S. Vickers, Entailment systems for stably locally compact locales, Theoretical Computer Science 316 (2004) 259--296.

\bibitem{b17} A.~Jung, M.~Kegelmann,  M.~A. Moshier, Multi lingual sequent calculus and coherent spaces, Fundamenta Informaticae,  37,  (1999) 369--412.


\bibitem{b18} A.~Jung, Continuous domain theory in logical form,Lecture Notes in Computer Science,  7860, (2013) 166--177.

\bibitem{b19} M. Bonsangue, J. Kok,
 Toward an infinitary logic of domains: Abramsky logic for transition systems,
Information and Computation, 155 (1999) 170--201.

\bibitem{b20}  L. Wang, Q. Li,  A representation of proper BC domains based on conjunctive sequent calculi,  Mathematical Structures in Computer Science, 155 (2020) 1--13.


\bibitem{wang2}
L.~Wang, Q.~Li,
A logic for Lawson compact algebraic L-domains,
 Theoretical Computer Science,  813 (2020) 410--427.

\bibitem{b21}
Y.~Chen, A.~Jung,
 A logical approach to stable Domains,
 Theoretical Computer Science 368 (2006) 124--148.


\bibitem{b22}  B.~A. Davey and H.~A. Priestley, Introduction to Lattices and Order, Cambridge,  UK: Cambridge University Press, 2002.


\bibitem{b23}  J.~Goubault-Larrecq, Non-Hausdorff Topology and Domain Theory, Cambridge,  UK: Cambridge University Press, 2013.

\bibitem{b24}  G.~Gierz, K.~H. Hofmann, K.~Keimel, J.~D. Lawson, M.~Mislove, D.~S. Scott, Continuous Lattices and Domains, Cambridge,  UK: Cambridge University Press, 2003.

\bibitem{b25} Y.~Chen,
Stone duality and
representation of stable domain,
Computers Math. Applic,  34 (1997) 27--41.

\bibitem{b26} G. Zhang,
 Disjunctive Systems and {L}-domains,
 Lecture Notes in
  Computer Science,  623 (1992) 284--295.

\bibitem{b27}  S.~Awodey, Category Theory, Oxford,  UK: Oxford University Press, 2006.





\end{thebibliography}
\end{document}